\documentstyle[epsf,float,seceq]{ptptex}

\title{%
First-Order Phase Transitions in Frustrated Spin Systems
}

\author{
Akihisa {\sc Koga}, Akira {\sc Kawaguchi}, Kouichi {\sc Okunishi}$^{*}$ 
and Norio Kawakami
}

\inst{
Department of Applied Physics, Osaka University, Osaka 565-0871\\
$^*$
Department of Physics, Niigata University, Igarashi 2, 
Niigata 950-2181}

\recdate{
}

\abst{%
We give a short review of our recent works on the
first-order quantum phase transitions 
in frustrated spin chains with orthogonal-dimer
structure.  When the ratio of the competing antiferromagnetic 
exchange couplings is varied, 
a first-order transition occurs between the dimer phase and the 
plaquette phase, which is accompanied by the discontinuity in
the spin excitation gap.  We further show that strong frustration 
triggers the phase transitions in a magnetic field, which exhibit 
plateaus and jumps  in the magnetization curve. 
The hole-doping effect is also addressed for the orthogonal-dimer
 chain with linked-tetrahedra structure.
It is found that the competing antiferromagnetic interactions
  result in a first-order metal-insulator transition upon hole doping.
}

\begin{document}

\maketitle

\section{Introduction}
Recently antiferromagnetic 
quantum spin systems with strong frustration have been 
studied intensively.
A typical example is  $\rm SrCu_2(BO_3)_2$\cite{Kageyama}
where the spin-1/2 magnetic Cu ions are located on the 
 orthogonal-dimer lattice (the so-called Shastry-Sutherland 
model).\cite{Shastry}
Various interesting phenomena were observed for this compound
such as dispersionless triplet excitations, 
correlated hopping of two magnons,  plateaus
in the magnetization curve.  It is suggested 
that novel properties in this compound are closely related to
frustration due to the competing 
antiferromagnetic couplings.\cite{Miyahara,SrTheory}  Besides such 
spin systems, metallic systems with frustration 
have also attracted much attention recently.
For instance, it is suggested that heavy-fermion behavior
observed  in $\rm LiV_2O_4$ \cite{Kondo} may be  caused by
geometrical frustration inherent in the pyrochlore-lattice 
structure.\cite{Kaps}

One of the remarkable properties common in such frustrated systems
is that strong frustration enhances the competition among
several eigenstates with distinct character as a candidate for the 
ground state, giving rise to first-order quantum phase transitions
when we vary the exchange couplings, the magnetic field,
the chemical potential, etc.
In order to address the role of geometrical frustration 
in quantum phase transitions, in this paper, we investigate
a simple one-dimensional (1D) version of the
frustrated spin model with orthogonal-dimer structure.
 By means of the series expansion technique,\cite{Singh} 
the exact diagonalization (ED) and 
the density matrix renormalization group (DMRG),\cite{White,pwfrg} 
we demonstrate that strong frustration indeed causes a 
wide variety of first-order quantum phase transitions.

\section{Quantum spin system with orthogonal-dimer structure}
In this section, we study first-order transitions 
in the 1D orthogonal-dimer spin chain.\cite{Ivanov,plachn} 
\begin{figure}[htb]
\epsfxsize=8cm
\centerline{\epsfbox{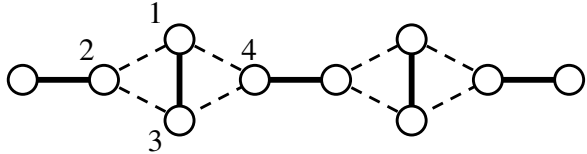}} 
\caption{Heisenberg spin chain with orthogonal-dimer structure. 
The solid (dashed) line represents the antiferromagnetic
coupling $J (J')$.}
\label{fig:plachn}
\end{figure}
The Hamiltonian we shall deal with
is  the standard  $S=1/2$ antiferromagnetic Heisenberg 
model in a magnetic field $H$, ${\cal H}=\sum J_{ij}{\bf S}_{i}
\cdot{\bf S}_{j}-H\sum S_{i}^{z}$, on the orthogonal-dimer
chain shown in Fig. \ref{fig:plachn}.
We will use the normalized parameters $j=J'/J$ and $h=H/J$ 
for convenience.

We first summarize the ground state properties of the orthogonal-dimer 
spin chain.\cite{Ivanov,plachn}
The remarkable point characteristic of the orthogonal-dimer system is that 
a direct product of local dimer-singlet states indicated by
 the solid line in Fig. 
\ref{fig:plachn} is always an exact eigenstate of the Hamiltonian.
Therefore, the dimer-singlet state should be the exact ground state up to 
a certain critical value of $j$.\cite{Ivanov}
\begin{figure}[htb]
\epsfxsize=11cm
\centerline{\epsfbox{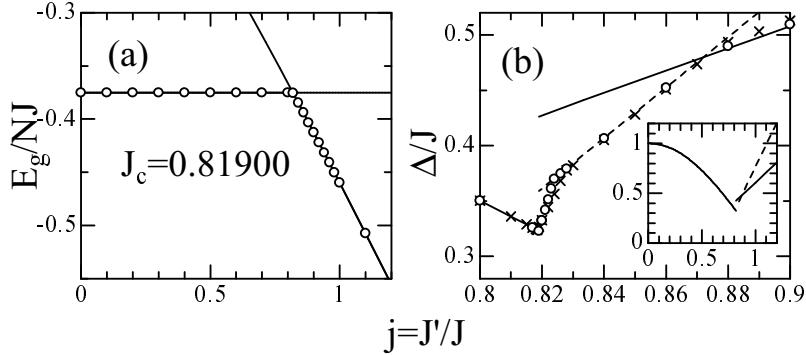}} 
\caption{(a) Ground state energy per site and
(b) magnetic excitation energy $\Delta$
as a function of $j$ for the orthogonal-dimer chain. 
Crosses and open circles
represent the results obtained  for the finite 
chain of $N=16, 24$ with periodic boundary conditions.
}
\label{fig:eg}
\end{figure}
On the other hand, in the large $j$ region, another singlet state
is stabilized, which 
is adiabatically connected to the isolated plaquette-singlet
denoted by the broken line in Fig. \ref{fig:plachn}.\cite{Kato,Koga1} 
As seen from Fig. \ref{fig:eg} (a),
a first-order quantum phase transition occurs  between the dimer and 
the plaquette phases at the critical point, $j_{c}=0.81900$.

Let us now turn to the spin excitation spectrum.
In the dimer phase  $(j<j_c)$, the orthogonal-dimer structure forbids
free hopping of a triplet excitation created over the local
dimer state.\cite{Shastry,Miyahara}
On the other hand, in the plaquette phase $(j>j_c)$ 
there are two kinds of low-energy magnetic excitations.
One is a dispersive triplet excitation, which is formed by 
simply breaking the plaquette singlet state. The corresponding  
excitation energy is shown as the solid line in Fig. \ref{fig:eg} (b).
The other is an unusual four-fold degenerate excitation 
shown as the dashed line, which is formed by breaking the
plaquette singlet state first, and then making a local dimer state
accompanied by 
two free spins.\cite{plachn}
Since the latter excitation becomes the lowest-excitation in the
region  $j_c<j<0.87$, as shown in Fig. \ref{fig:eg},
it plays an essential role for the phase transitions
in the vicinity of the critical point, as will be shown below.  
In this way, the unusual four-fold degenerate excitation found here
possesses intermediate properties 
between those typical for the dimer phase and the plaquette phase.

\begin{figure}[htb]
\epsfxsize=11cm
\centerline{\epsfbox{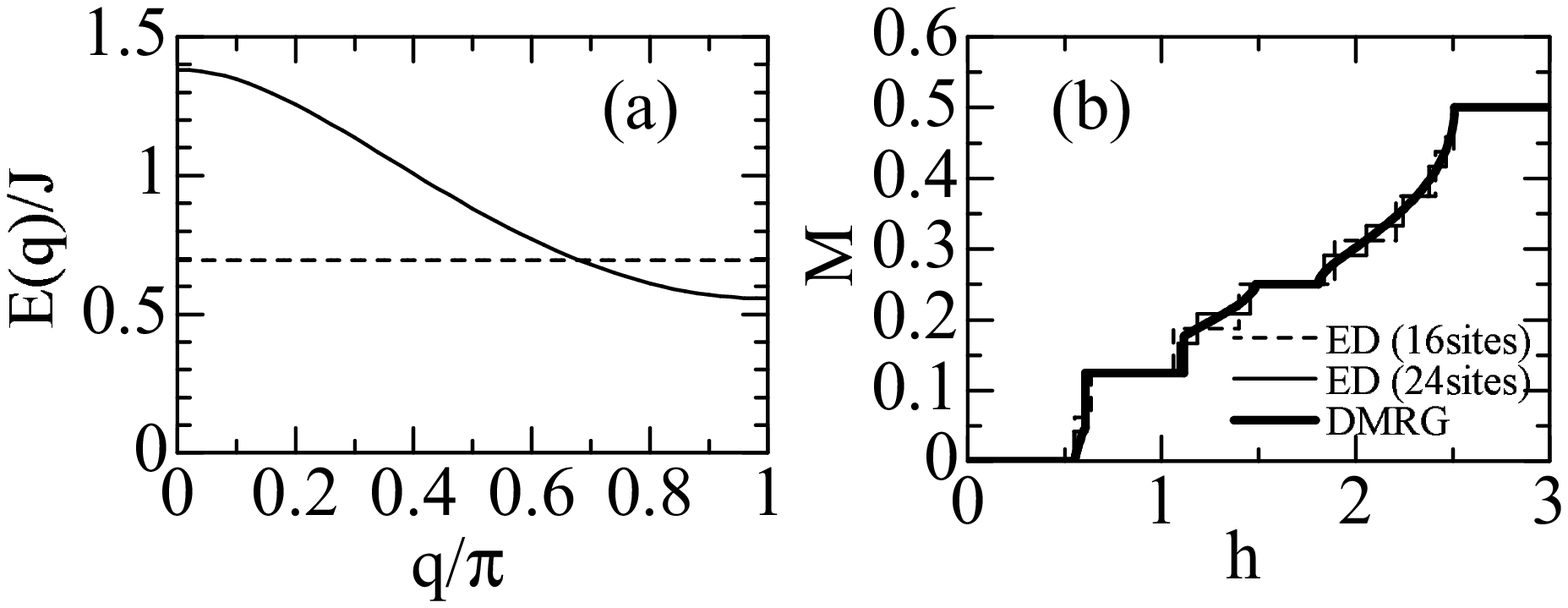}} 
\caption{(a) Dispersion relation calculate by 
the series expansion method \cite{plachn} and (b) magnetization curve
calculated by ED and DMRG for $j=0.94$. 
}
\label{fig:dis94}
\end{figure}

We have seen so far that several different kinds of low-energy excitations 
coexist in our frustrated model. It
is  thus interesting to see how these excitations affect 
the magnetization process.  To this end, 
we focus here on the magnetization curve for the ratio of 
the exchange couplings $j=0.94$, where the system belongs 
to the plaquette phase. As shown in Fig. \ref{fig:dis94},
the lowest excitation is a plaquette-triplet excitation (solid line),
 and another four-fold dispersionless excitation (broken line)
lies slightly above it.
The coexistence of two-kinds of distinct excitations influences 
the magnetization curve considerably.
As $h$ increases beyond the critical field $(h_c=\Delta)$, 
the magnetization should develop with $M\sim (h-h_c)^{1/2}$, 
since it is dominated by the triplet excitation whose 
dispersion relation is quadratic near the bottom $(q=\pi)$.
With slight increase in $h$, however, the magnetization 
suddenly jumps to the $1/4$-plateau.
This first-order transition results from the coexistence of 
two distinct levels in low-energy excitations.  Therefore, 
it is instructive to remark
that the 1/4-plateau is not generated by 
the crystallization of the lowest triplet excitation, but 
of the four-fold degenerate excitation.
Beyond the 1/4-plateau, we again encounter 
another  first-order phase transition 
accompanied by a jump in $M$.

As seen in this section,  
nontrivial behavior shows up reflecting 
the competing interactions in the present model.
We wish to observe below what happens for such 1D spin systems, when 
we introduce the charge degree of freedom, which is important to discuss
a metallic system with strong frustration.

\section{Correlated electron system with strong frustration}

\begin{figure}[htb]
\epsfxsize=8cm
\centerline{\epsfbox{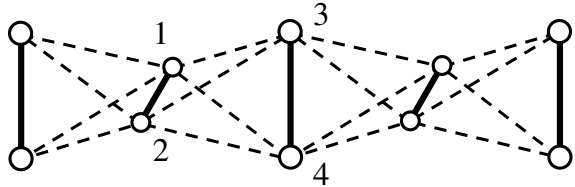}} 
\caption{Orthogonal-dimer spin chain with liked-tetrahedra structure.
The solid and the dashed lines correspond to the parameters $J (t)$ and 
$J' (t')$, respectively.
}
\label{fig:ladder}
\end{figure}

In this section, we extend our discussions to a hole-doped  
system.\cite{Ogata}
We here employ another orthogonal-dimer model \cite{Gellad}
 shown in Fig. \ref{fig:ladder}.
A nice feature in this model is that it
can describe not only the orthogonal-dimer spin system
but also the linked-tetrahedra system, which plays an important role 
for  understanding low-energy properties of the pyrochlore system.
In the undoped case, a first-order quantum phase transition 
between the dimer phase and the plaquette phase 
occurs at the critical point $j_c=0.71$.\cite{Gellad,Honecker}

We shall deal with a hole-doped system by introducing 
the {\it t-J} model with electron hopping, $t$ and $t'$, 
and clarify how 
frustration affects the nature of the metal-insulator transition.
Let us  first consider the hole-doping effects on the dimer phase. 
For simplicity, we set $t/J=t'/J'=4.0$. 
\begin{figure}[htb]
\epsfxsize=11cm
\centerline{\epsfbox{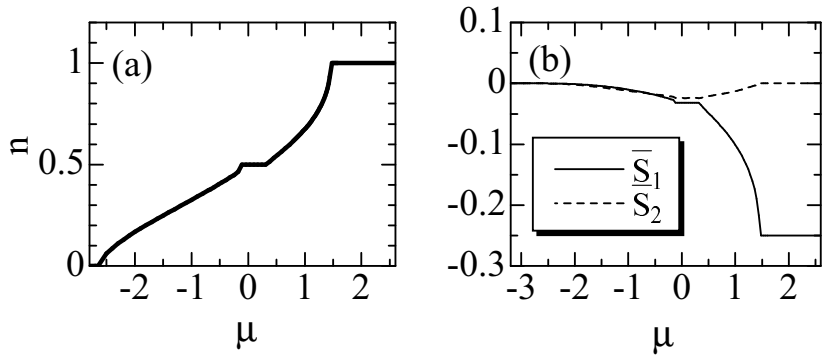}} 
\caption{(a) Electron density $n$ and 
(b) short-range correlation $\bar{S}_\alpha$
as a function of the chemical potential $\mu$ for 
$j=0.4$ (dimer phase).
}
\label{mu_n_04}
\end{figure}
In Fig. \ref{mu_n_04} (a), the electron density $n$ calculated by DMRG
is shown as a function of 
the chemical potential $\mu$ for $j=0.4$. 
In this figure, $n=1$ corresponds to half-filling at which 
the system is in the insulating phase with singlet-dimer state. 
It is seen that hole-doping smoothly 
drives the system to a metallic phase. 
At quarter filling ($n=1/2$), we encounter another insulating phase, 
which is stabilized by the formation of a CDW state.


In order to see the nature of spin correlations in the metallic phase,  
we compute the  short-range spin correlation functions
in the rung and chain directions, which are defined as
(see Fig. \ref{fig:ladder}),
\begin{eqnarray}
\bar{S}_1&=&
    \frac{1}{2}
    ( <S^z_1 S^z_2>+<S^z_3 S^z_4> ), 
\nonumber \\
\bar{S}_2  &=&
     \frac{1}{4}
     ( <S^z_1 S^z_3>+<S^z_1 S^z_4> 
     +<S^z_2 S^z_3>+<S^z_2 S^z_4>   ). 
\nonumber
\label{TJ_EQ}
\end{eqnarray}
In Fig. \ref{mu_n_04} (b), the spin correlation functions 
calculated by DMRG are shown
as a function of the chemical potential $\mu$.  Note that we have 
$\bar{S}_1=-1/4$ and $\bar{S}_2=0$ at half filling, which accord with 
those expected  for  the  isolated
dimers. It is seen that the spin correlation along the 
rung, $\bar{S}_1$, rapidly decreases down to $-1/4$
as the system approaches half filling from the metallic side.  
On the other hand, $\bar{S}_2$ becomes almost zero
along the chain direction, reflecting the fact that the 
 system is an assembly of decoupled dimers at half filling.
Therefore, near half filling, the metallic state is considered 
as a resonating state composed of such dimer pairs.

In contrast to the dimer phase, strong frustration dramatically changes 
 the nature of the metal-insulator transition
in the plaquette phase.
\begin{figure}[htb]
\epsfxsize=11cm
\centerline{\epsfbox{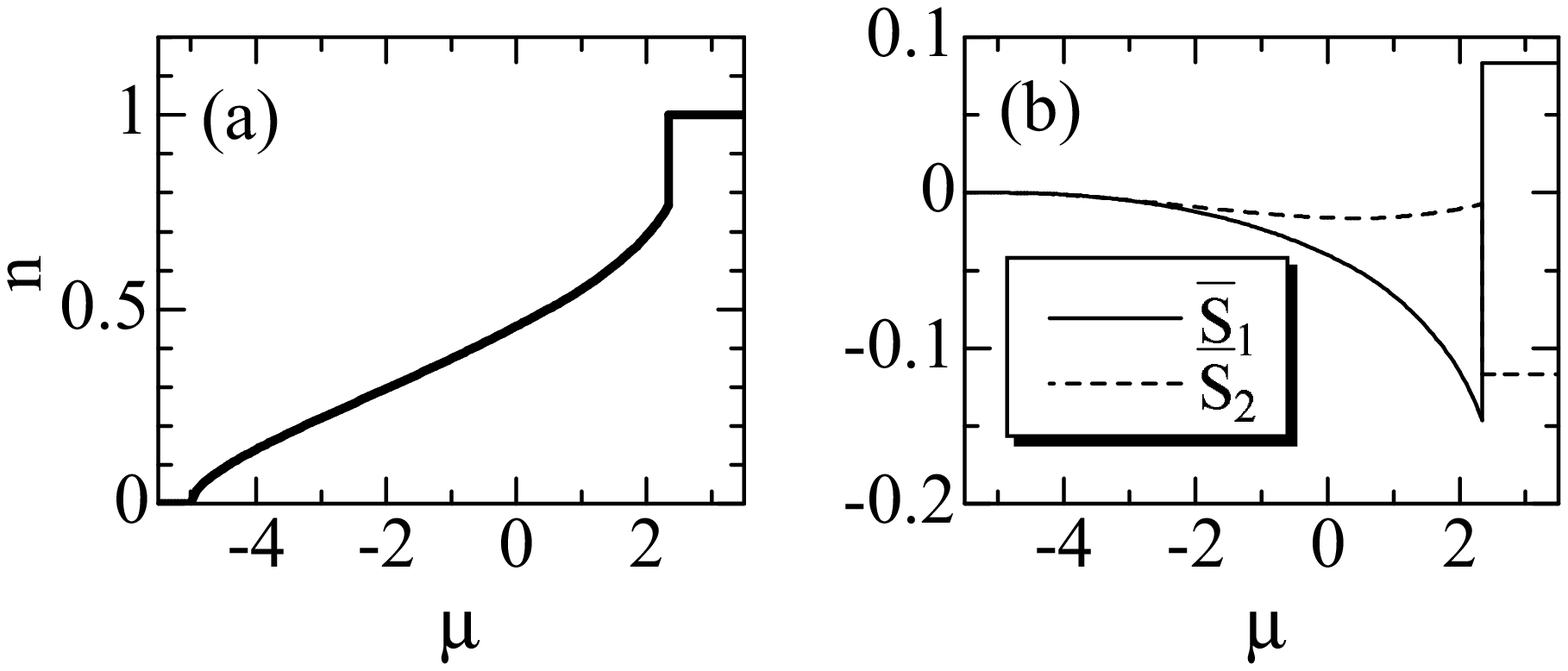}} 
\caption{(a) Electron density $n$  and 
(b) short-range correlation $\bar{S}_\alpha$
as a function of $\mu$ for $j=1.0$ (plaquette phase).
}
\label{mu_n_10}
\end{figure}
In Fig. \ref{mu_n_10} (a), we show the electron 
density $n$ as a function of $\mu$
for $j=1.0$, where the system is in the plaquette-singlet phase at 
half filling.  It should be noticed that 
as $\mu$ decreases from half filling ($n=1$), a
first-order phase transition 
suddenly occurs to the metallic phase at $\mu\simeq 2.2$,
which is accompanied by the discontinuity in the 
electron  density.
To clarify  the nature of the metal-insulator transition 
in more detail, we again discuss
the short-range spin correlation function
[see Fig. \ref{mu_n_10} (b)].
In the insulating phase ($n=1$) for  $\mu > 2.2$, the ground state is 
in the plaquette-singlet phase, which is characterized by 
$\bar{S}_1>0$ and $\bar{S}_2<0$ in accordance with the numerical results.  
On the other hand,  once the first-order transition
occurs, both of 
$\bar{S}_1$ and $\bar{S}_2$ have negative values in the metallic phase.
Note that $\bar{S}_1$ decreases with negative values 
while $\bar{S}_2$ gets very small
as the system approaches half-filling from the metallic side.  
This implies that 
the dimer correlation along the rung is largely enhanced
in the metallic phase close to half filling.
This dimer-dominant state competes
with the plaquette-singlet state realized at half filling, triggering
 the first-order phase transition.




\section{Summary}
We have investigated the  first-order quantum phase transitions in 
antiferromagnetic spin chains with orthogonal-dimer structure.
It has been clarified that there appears rich structure 
in the excitation spectrum, which is caused by strong 
 frustration. Such nontrivial excitations determine the nature of 
the quantum phase transitions, when we change the 
interactions, the magnetic field, etc. We have 
also studied the hole-doping effect on another 
orthogonal-dimer chain with linked-tetrahedra 
structure. It has been
shown that the first-order metal-insulator transition 
is triggered by strong geometrical frustration.  
We have checked in a preliminary 
calculation that such a first-order metal-insulator transition
also appears in the model of \S 2, implying that
it may be common in this class of the 
 orthogonal-dimer models.

\section*{Acknowledgements}
This work was partly supported by a Grant-in-Aid from the Ministry 
of Education, Science, Sports and Culture of Japan. 
A part of computations was done at the Supercomputer Center at the 
Institute for Solid State Physics, University of Tokyo
and Yukawa Institute Computer Facility. 
A. Kawaguchi is supported by the Japan Society 
for the Promotion of Science.


\begin{thebibliography}{99}

\bibitem{Kageyama}
H. Kageyama, K. Yoshimura, R. Stern, N. V. Mushnikov, K. Onizuka, M. Kato,
K. Kosuge, C. P. Slichter, T. Goto and Y. Ueda:
Phys. Rev. Lett. {\bf 82} 3168 (1999).

\bibitem{Shastry}
B. S. Shastry and B. Sutherland: Physica {\bf 108B} 1069 (1981).

\bibitem{Miyahara}
S. Miyahara and K. Ueda: Phys. Rev. Lett. {\bf 82}, 3701 (1999).

\bibitem{SrTheory}
A. Koga and N. Kawakami: Phys. Rev. Lett. {\bf 84}, 4461 (2000);
C. Knetter, A. B\"uhlet, E. M\"uller-Hartmann and G. S. Uhrig:
Phys. Rev. Lett. {\bf 85}, 3958 (2000);
T. Momoi and K. Totsuka, Phys. Rev. B {\bf 61} 3231 (2000);
Y. Fukumoto: J. Phys. Soc. Jpn. {\bf 69}, 2755 (2000);
G. Misguish, Th. Jolicoeur and S. M. Girvin, 
Phys. Rev. Lett. {\bf 87} 097203 (2001);
C. H. Chung, J. B. Marston and S. Sachdev, Phys. Rev. B {\bf 64} 134407 (2001).


\bibitem{Kondo}
S. Kondo, D. C. Johnston, C. A. Swenson, F. Borsa, A. V. Mahajan, L. L. Miller,
T. Gu, A. I. Goldman, M. B. Maple, D. A. Gajewski, E. J. Freeman, N. R. Dilley,
R. P. Dickey, J. Merrin, K. Kojima, G. M. Luke, Y. J. Uemura, O. Chmaissem 
and J. D. Jorgensen, Phys. Rev. Lett. {\bf 78}, 3729 (1997).

\bibitem{Kaps}
H. Kaps, N. B\"uttgen, W. Trinkl, A. Loidl, M. Klemm and S. Horn,
J. Phys. Condens. Matter {\bf 13}, 8497 (2001).

\bibitem{Singh}
M. P. Gelfand and R. R. P. Singh, Adv. Phys. {\bf 49}, 93 (2000).

\bibitem{White}
S. R. White, Phys. Rev. Lett. {\bf 69}, 2863 (1992);
Phys. Rev. B {\bf 48}, 10345 (1993).

\bibitem{pwfrg} 
T. Nishino and K. Okunishi, 
J. Phys. Soc. Jpn. {\bf 63}, 4084 (1995);
Y. Hieida, K. Okunishi and Y. Akutsu, 
Phys. Lett. A {\bf 233}, 464 (1997).

\bibitem{Ivanov}
N. B. Ivanov and J. Richter,
Phys. Lett. {\bf 232A}, 308 (1997);
J. Richter, N. B. Ivanov and J. Schulenburg,
J. Phys. Condence Matt. {\bf 10}, 3635 (1998);
J. Schulenburg and J. Richter, 
Phys. Rev. B {\bf 65}, 054420 (2002).

\bibitem{plachn}
A. Koga, K. Okunishi and N. Kawakami,
Phys. Rev. B {\bf 62}, 5558 (2000).

\bibitem{Kato}
N. Kato and M. Imada,
J. Phys. Soc. Jpn. {\bf 64}, (1995) 4105.

\bibitem{Koga1}
A. Koga, S. Kumada, N. Kawakami and T. Fukui, 
J. Phys. Soc. Jpn. {\bf 67}, 622 (1998).

\bibitem{Ogata}
M. Ogata, M. U. Luchini, T. M. Rice,
Phys. Rev. B {\bf 44} (1991)  12083.

\bibitem{Gellad}
M. P. Gelfand, Phys. Rev. B {\bf 43}, 8644 (1991).

\bibitem{Honecker}
A. Honecker, F. Mila and M. Troyer, Eur. Phys. J. B, {\bf 15} 227,(2000).









\end{thebibliography}
\end{document}